\documentclass{PoS}

\title{An MCMC study of general squark flavour mixing in the MSSM}
\ShortTitle{An MCMC study of general squark flavour mixing in the MSSM}

\author{\speaker{Bj\"orn Herrmann}\\
  LAPTh, Universit\'e Savoie Mont Blanc, CNRS, 9 Chemin de Bellevue, F-74941 Annecy-le-Vieux, France\\
  E-mail: \email{herrmann@lapth.cnrs.fr}}

\author{Karen De Causmaecker\\
  Theoretische Natuurkunde, IIHE/ELEM and International Solvay Institutes, Vrije Universiteit Brussel, Pleinlaan 2, B-1050 Brussels, Belgium}

\author{Benjamin Fuks\\
  Sorbonne Universit\'es, UPMC Univ.\ Paris 06, UMR 7589, LPTHE, F-75005 Paris, France;\\
  CNRS, UMR 7589, LPTHE, F-75005 Paris, France}

\author{Farvah Mahmoudi\\
  Universit{\'e} de Lyon, Universit{\'e} Lyon 1, F-69622 Villeurbanne Cedex, France;
    Centre de Recherche Astrophysique de Lyon, CNRS, UMR 5574, F-69561 Saint-Genis Laval Cedex, France;
    Ecole Normale Sup{\' e}rieure de Lyon, France; \\
	CERN Theory Division, Physics Department, CH-1211 Geneva 23, Switzerland; \\
	Institut Universitaire de France, 103 boulevard Saint-Michel, 75005 Paris, France}

\author{Ben O'Leary, Werner Porod\\
  Institut f\"ur Theoretische Physik und Astrophysik, Universit\"at W\"urzburg, D-97074 W\"urzburg, Germany}

\author{Sezen Sekmen\\
  Kyungpook National University, Daegu, 702-701, South Korea}

\author{Nadja Strobbe\\
  Fermi National Accelerator Laboratory, Batavia, 60510-5011, USA}

\abstract{We present an extensive study of non-minimally flavour violating (NMFV) terms in the Lagrangian of the Minimal Supersymmetric Standard Model (MSSM). We impose a variety of theoretical and experimental constraints and perform a detailed scan of the parameter space by means of a Markov Chain Monte-Carlo (MCMC) setup. This represents the first study of several non-zero flavour-violating elements within the MSSM. We present the results of the MCMC scan with a special focus on the flavour-violating parameters. Based on these results, we define benchmark scenarios for future studies of NMFV effects at the LHC.}

\FullConference{The European Physical Society Conference on High Energy Physics\\
		22--29 July 2015\\
		Vienna, Austria}

\begin{document}

\section{Introduction} 
\label{sec:intro}

Despite a large experimental search programme, no signal of Supersymmetry (SUSY) has been found so far at the Large Hadron Collider (LHC). Therefore, supersymmetric particles are constrained to reside at scales that are not reachable at LHC, or the spectrum must present specific properties that allow the superpartners to escape detection. Following the latter guiding principle, we consider the framework of non-minimal flavour violation (NMFV) in the squark sector, where flavour-violating terms, which are not related to the CKM matrix, are allowed in the mass matrices at the TeV scale. We explore, by means of a dedicated Markov Chain Monte Carlo analysis to which extent deviations from minimal flavour violation (MFV) are allowed by theoretical constraints and current experimental data. 

The impact of NMFV terms in the squark sector has already been investigated in various areas \cite{Buchalla:2008jp, Kowalska:2014opa}. All previous studies, however, have relied on allowing only few non-zero NMFV elements in the mass matrices. However, one would generally expect several of them to be non-vanishing, especially when the low-energy model is supposed to emerge from a larger framework such as Grand Unified Theories where the flavour structure is generated at some higher scale. We therefore undertake a comprehensive study of NMFV in the MSSM taking into account all flavour-violating terms involving the second and third generations of squarks. 

In the following, after introducing the model, notation, and setup, we present selected results and aspects emerging from this study. In particular, after introducing the model and parameters, we discuss the distributions of the quark-flavour conserving (QFC) and violating (QFV) parameters at the TeV scale together with selected observables such as squark masses and flavour decompositions. Based on our analysis, we present benchmark scenarios intended for future studies. For the complete analysis we refer the reader to Ref.\ \cite{OurPaper}.

\section{Squark mixing in the MSSM}
\label{sec:model}

In the super-CKM basis, the up-type squark sector is parametrized by the squared mass matrix
\begin{equation}
	M_{\tilde{u}}^2 ~=~ \left( \begin{array}{cc}
		V_{\rm CKM} M_{\tilde{Q}}^2 V_{\rm CKM}^{\dag} + m_u^2 + D_{u,L} & \frac{v_u}{\sqrt{2}} T_u^{\dag} - m_u \frac{\mu}{\tan\beta} \\
		\frac{v_u}{\sqrt{2}} T_u - m_u \frac{\mu}{\tan\beta} & M_{\tilde{U}}^2 + m_u^2 + D_{u,R}
		\end{array} \right)
\label{Eq:MassMatrix}
\end{equation}
which includes the soft SUSY breaking mass matrices $M_{\tilde{Q}}^2$ and $M_{\tilde{U}}^2$ as well as the trilinear coupling matrix $T_u$, which is related to the Yukawa matrix $Y_u$ through $(T_u)_{ii} = (A_u)_{ii} (Y_u)_{ii}$ ($i=1,2,3$). In the framework of MFV, these matrices are diagonal in flavour space, while in the context of NMFV off-diagonal elements may be present. The remaining parameters appearing in Eq.\ (\ref{Eq:MassMatrix}) are the Higgsino mass parameter $\mu$, the ratio of the vacuum expectation values of the two Higgs doublets $\tan\beta=v_u/v_d$, the CKM-matrix $V_{\rm CKM}$, and the up-quark mass matrix $m_u$. Finally, the $D$-terms contain only SM parameters together with the angle $\beta$ and are diagonal in flavour space. A similar expression holds for the mass matrix of the down-type squarks.

In our study, we assume that the first two generations are mass-degenerate, such that we have $(M_{\tilde{Q}})_{11} = (M_{\tilde{Q}})_{22} \equiv M_{\tilde{Q}_{1,2}}$, $(M_{\tilde{U}})_{11} = (M_{\tilde{U}})_{22} \equiv M_{\tilde{U}_{1,2}}$, and $(M_{\tilde{D}})_{11} = (M_{\tilde{D}})_{22} \equiv M_{\tilde{D}_{1,2}}$. For the third-generation mass parameters we adopt the simplified notation $(M_{\tilde{Q}})_{33} \equiv M_{\tilde{Q}_3}$, $(M_{\tilde{U}})_{33} \equiv M_{\tilde{U}_3}$, and $(M_{\tilde{D}})_{33} \equiv M_{\tilde{D}_3}$. Moreover, we assume that only the third-generation trilinear couplings are non-zero and we take them equal for simplicity, $(A_u)_{33} = (A_d)_{33} = (A_{\ell})_{33} \equiv A_f$. 

Restricting ourselves to mixing between the second and third generations \cite{Ciuchini:2007ha}, seven independent NMFV elements are present in the mass matrices at the TeV scale. We parametrize them in a dimensionless and scenario-independent manner according to 
\begin{eqnarray}
	\delta_{LL} = \frac{(M_{\tilde{Q}}^2)_{23}}{ M_{\tilde{Q}_{1,2}} M_{\tilde{Q}_3} },
	\quad
	\delta_{RR}^u = \frac{(M_{\tilde{U}}^2)_{23}}{ M_{\tilde{U}_{1,2}} M_{\tilde{U}_3} },
	\quad
	\delta_{RL}^u = \frac{v_u}{\sqrt{2}}\frac{(T_u)_{23}}{ M_{\tilde{Q}_{1,2}} M_{\tilde{U}_3} }, 
	\quad
	\delta_{LR}^u = \frac{v_u}{\sqrt{2}}\frac{(T_u)_{32}}{ M_{\tilde{Q}_3} M_{\tilde{U}_{1,2}} },
	~~~~~
	\label{Eq:NMFVParams}
\end{eqnarray}
and three corresponding expressions defining the parameters $\delta^d_{RR}$, $\delta^d_{LR}$ and $\delta^d_{RL}$ related to the down-type squarks. 

In the Higgs sector, the independent parameters are $\mu$, $\tan\beta$, and the pole mass $m_A$ of the pseudoscalar Higgs boson. The model parametrization is completed by the common slepton mass parameter $M_{\ell}$ and the bino mass parameter which we relate to the wino and gluino masses through the ``GUT relation'' $M_1 = M_2/2 = M_3/6$. All parameters that are considered in our study and their ranges are summarized in Table \ref{TabParams}.

Our goal is to find and characterize regions in this parameter space which are consistent with constraints from flavour physics, Higgs boson mass and anomalous magnetic moment of the muon, and which also have the lightest supersymmetric particle (LSP) to be the lightest neutralino, in order to have a viable dark matter candidate. To do so, we sample points from the parameter space using the Markov Chain Monte Carlo method \cite{Markov} based on a likelihood composed of the above constraints, a complete list of which is given in Table \ref{TabConstraints}. For each parameter point, the physical mass spectrum is computed using the public programme {\tt SPheno} \cite{Porod:2003um}, and the numerical evaluation of the relevant observables has been done using {\tt SPheno} and {\tt SuperIso} \cite{Mahmoudi:2007vz} relying on the SUSY \cite{Allanach:2008qq} and Flavour \cite{Mahmoudi:2010iz} Les Houches Accords. 

\begin{table}
\renewcommand{\arraystretch}{1.2}
  \begin{center}   	  
	  \begin{picture}(500,170)
	  	\put(5,125){
	  		\begin{tabular}{|c|c|}
	  			\hline
        		~SM Parameter ~& Scanned range \\
     		    \hline
        		$\alpha_s(m_Z)$ & $N(0.1184,0.0007)$ \\
        		$m_t^{\rm pole}$ & $N(173.3,1.3928)$ \\
        		$m_b(m_b)$ & $N(4.19,0.12)$ \\
      		  	\hline
			\end{tabular}				
		}
	  	\put(5,40){
		    \begin{tabular}{|c|c|}
		      \hline
		        QFV Parameter & ~~~ Scanned range ~~~\\
		      \hline
		        $\delta_{LL}$, $\delta_{RR}^u$, $\delta_{RR}^d$ & [-0.8, 0.8] \\
		        $\delta_{LR}^u$, $\delta_{RL}^u$ & [-0.5, 0.5] \\
		        $\delta_{LR}^d$, $\delta_{RL}^d$ & [-0.05, 0.05] \\
			  \hline
		    \end{tabular}
		}
	  	\put(205,88){
		    \begin{tabular}{|c|c|}
			  \hline
		        QFC Parameter & Scanned range \\
		      \hline
		        $M_{\tilde{Q}_{1,2}}$, $M_{\tilde{U}_{1,2}}$, $M_{\tilde{D}_{1,2}}$ & [300, 3500] \\
		        $M_{\tilde{Q}_{3}}$, $M_{\tilde{U}_{3}}$, $M_{\tilde{D}_{3}}$, $M_{\tilde{\ell}}$ & [100, 3500] \\
				\hline
		        $M_1$, $m_{A}$ & [100, 1600] \\
		        $\tan \beta$ & [10, 50] \\
		        $\mu$ & [100, 850] \\
		      \hline
		        $A_f$ & [-10000, 10000] \\
		           	  & {\rm or} $|A_f| < 4 \max\{M_{\tilde{q}}, M_{\tilde{\ell}}\}$ \\
			  \hline
		    \end{tabular}		
		}
	 \end{picture}
	\caption{Standard Model (SM), supersymmetric quark-flavour conserving (QFC) and violating (QFV) parameters of the NMFV MSSM parameter space. All dimensionful quantities are given in GeV. $N(\mu,\sigma)$ denotes a Gaussian profile of mean $\mu$ and width $\sigma$.}
	\label{TabParams}
\end{center}
\end{table}

\begin{table}
\renewcommand{\arraystretch}{1.2}
  \begin{center}
    \begin{tabular}{|c|c|c|}
      \hline
        Observable & Experimental result & Likelihood function \\
      \hline
        ${\rm BR}(B \rightarrow X_s\gamma) $ 
          & $(3.43 \pm 0.22) \times 10^{-4}$ \cite{Amhis:2014hma}
          & Gaussian\\
        ${\rm BR}(B_s \rightarrow \mu \mu)$
          & $(2.8 \pm 0.7)\times 10^{-9}$ \cite{CMS:2014xfa}
          & Gaussian \\
        ${\rm BR}(B \rightarrow K^* \mu \mu)_{q^2 \in [1,6] ~{\rm GeV}^2}$
          & $(1.7 \pm 0.31) \times 10^{-7}$ \cite{Aaij:2013iag}
          & Gaussian \\
        ${\rm AFB}(B \rightarrow K^* \mu \mu)_{q^2 \in [1.1,6] ~{\rm GeV}^2}$
          & $(-0.075 \pm 0.036) \times 10^{-7} $ \cite{LHCb:2015dla}
          & Gaussian \\
        ${\rm BR}(B \rightarrow X_s \mu \mu)_{q^2 \in [1,6] ~{\rm GeV}^2}$
          & $(0.66 \pm 0.88) \times 10^{-6}$ \cite{Lees:2013nxa}
          & Gaussian \\
        ${\rm BR}(B \rightarrow X_s \mu \mu)_{q^2 > 14.4 ~{\rm GeV}^2}$ 
          & $(0.60 \pm 0.31) \times 10^{-6}$ \cite{Lees:2013nxa}
          & Gaussian \\
        ${\rm BR}(B_u \rightarrow \tau \nu)/{\rm BR}(B_u \rightarrow \tau \nu)_{\rm SM} $
          & $1.04\pm 0.34$~\cite{Agashe:2014kda}
          & Gaussian \\
        $\Delta M_{B_s}$
          & $(17.719 \pm 3.300)\; {\rm ps}^{-1}$~\cite{Agashe:2014kda}
          & Gaussian \\
      \hline
        $\epsilon_K$
          & $(2.228 \pm 0.29^{\rm th}) \times 10^{-3}$~\cite{Agashe:2014kda}
          & Gaussian\\
        ${\rm BR}(K^0 \rightarrow \pi^0 \nu \nu)$
          & $  \leq 2.6 \times 10^{-8}$~\cite{Agashe:2014kda}
          & 1 if yes, 0 if no \\
        ${\rm BR}(K^+ \rightarrow \pi^+ \nu \nu)$
          & $ 1.73^{+1.15}_{-1.05} \times 10^{-10}$~\cite{Agashe:2014kda}
          & Two-sided Gaussian \\
      \hline
        $\Delta a_\mu$
          & $(26.1 \pm 12.8)\times 10^{-10}$ $[e^+e^-]$~\cite{Agashe:2014kda}
          & Gaussian \\
      \hline
        $m_h$
          & $125.5\pm2.5$ GeV~\cite{Aad:2012tfa} & 1 if yes, 0 if no \\
      \hline
        Lightest supersymmetric particle & Lightest neutralino & 1 if yes, 0 if no \\
	  \hline
    \end{tabular}
    \caption{Experimental constraints imposed in our scan of the NMFV MSSM parameter space.}
  \label{TabConstraints}
  \end{center}
\end{table}

\section{Selected results}
\label{sec:results}

In this section, we give a short overview of the most important results obtained from the Markov Chain Monte Carlo study described above. The full discussion, including additional details such as the influence of single constraints on the various NMFV parameters, can be found in Ref.\ \cite{OurPaper}.

\begin{figure}[t]
	\begin{center}
		\includegraphics[width=0.32\textwidth]{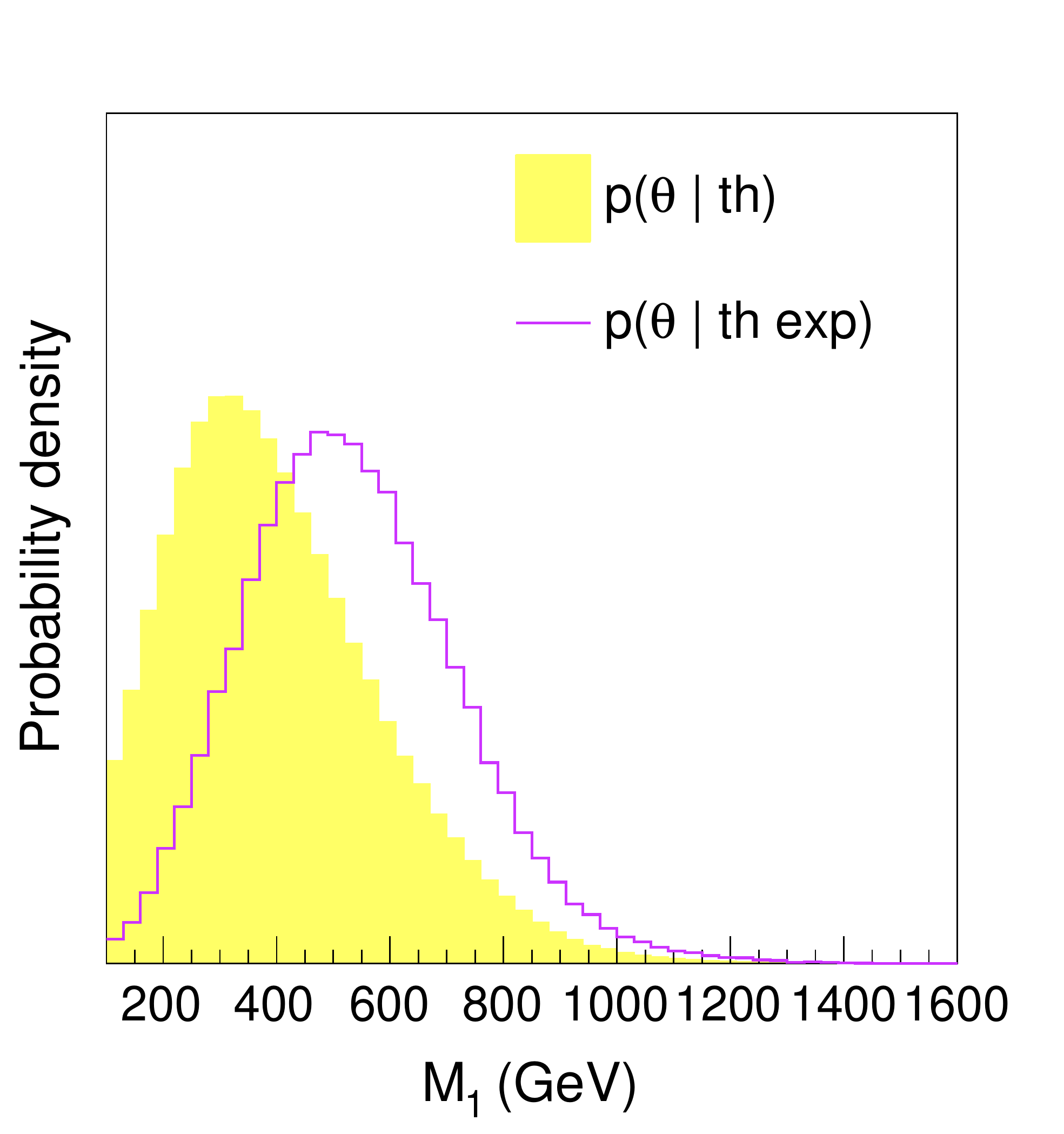} 
		\includegraphics[width=0.32\textwidth]{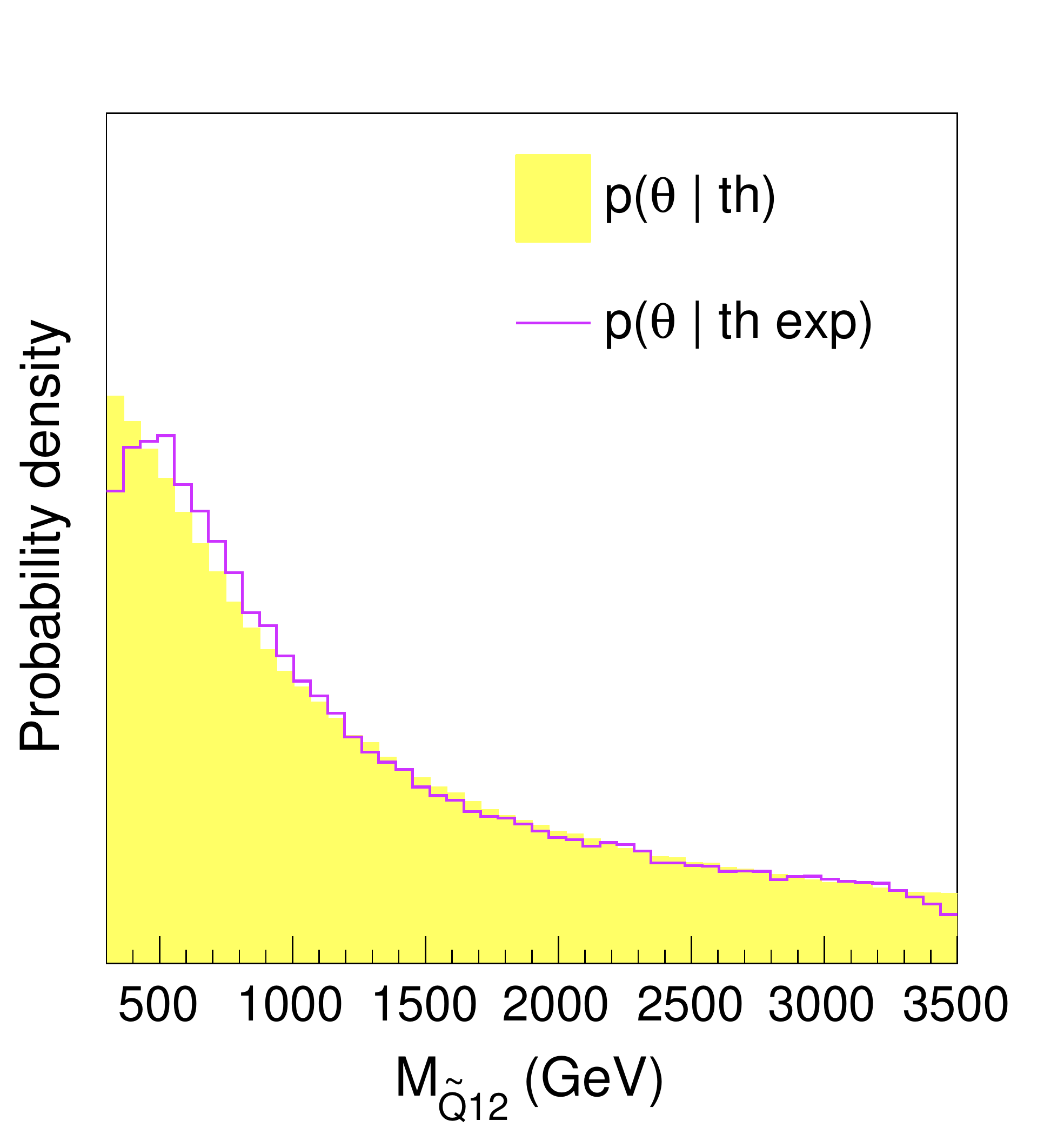} 
		\includegraphics[width=0.32\textwidth]{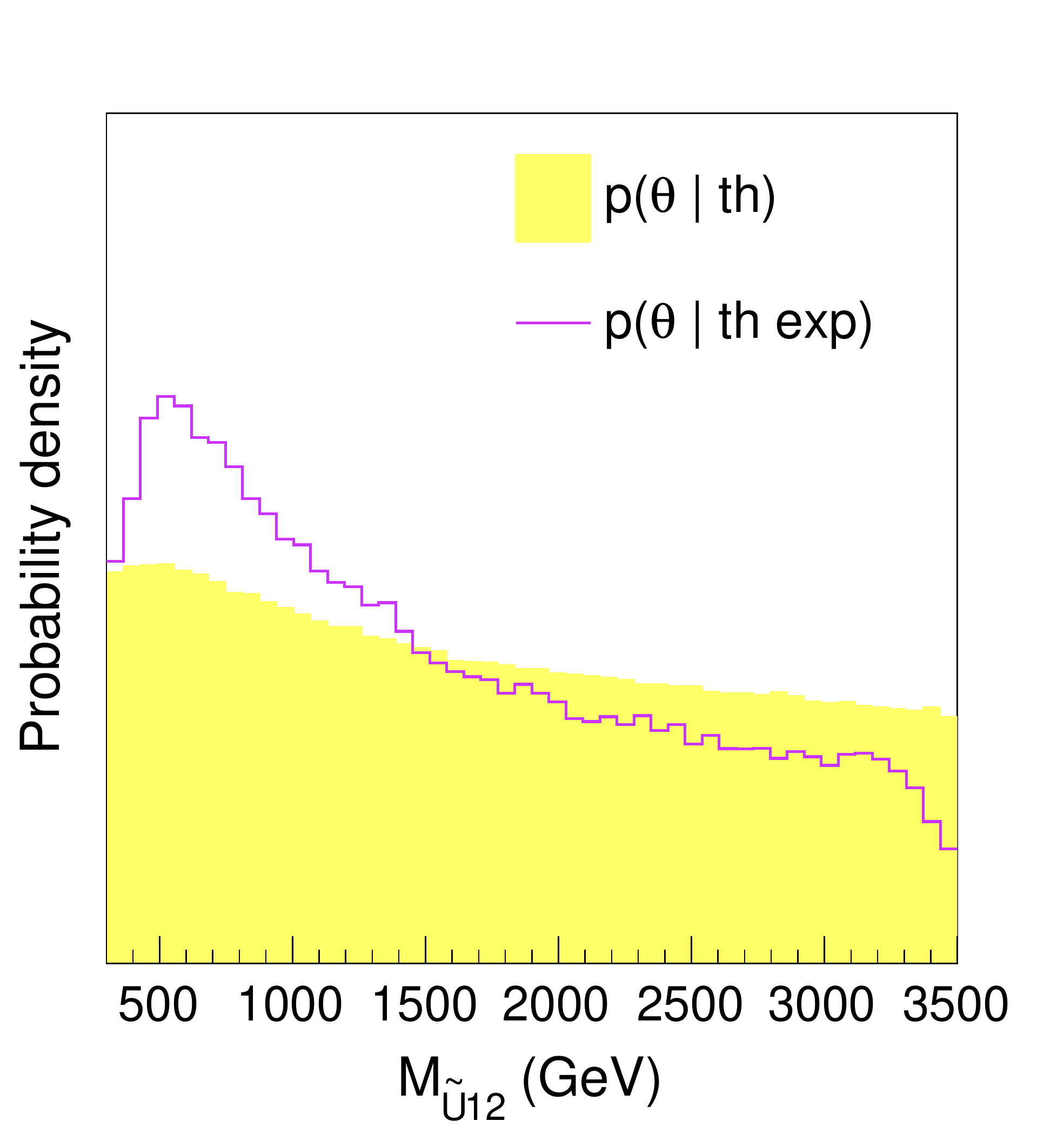} \\ 
		\includegraphics[width=0.32\textwidth]{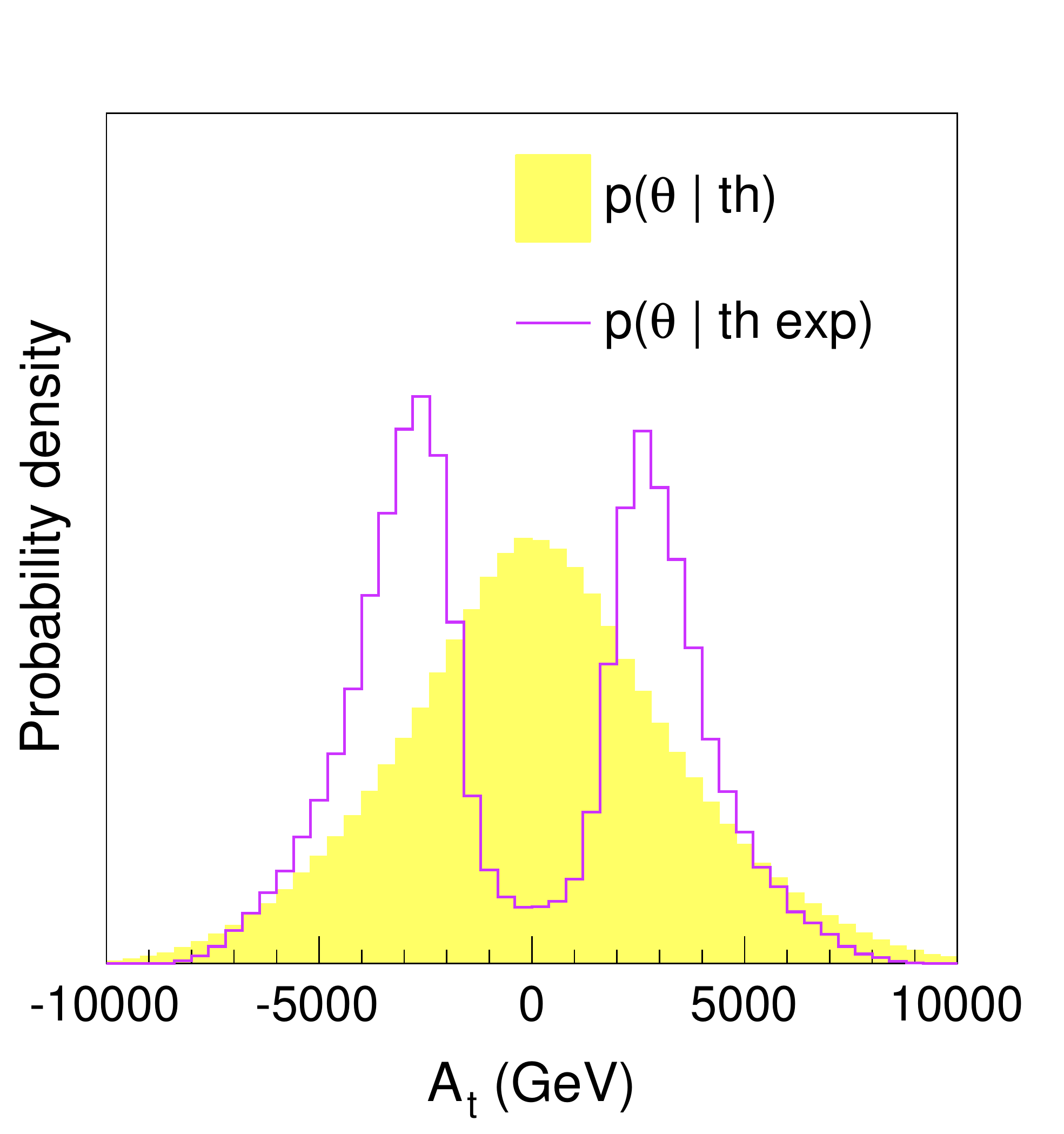} 
		\includegraphics[width=0.32\textwidth]{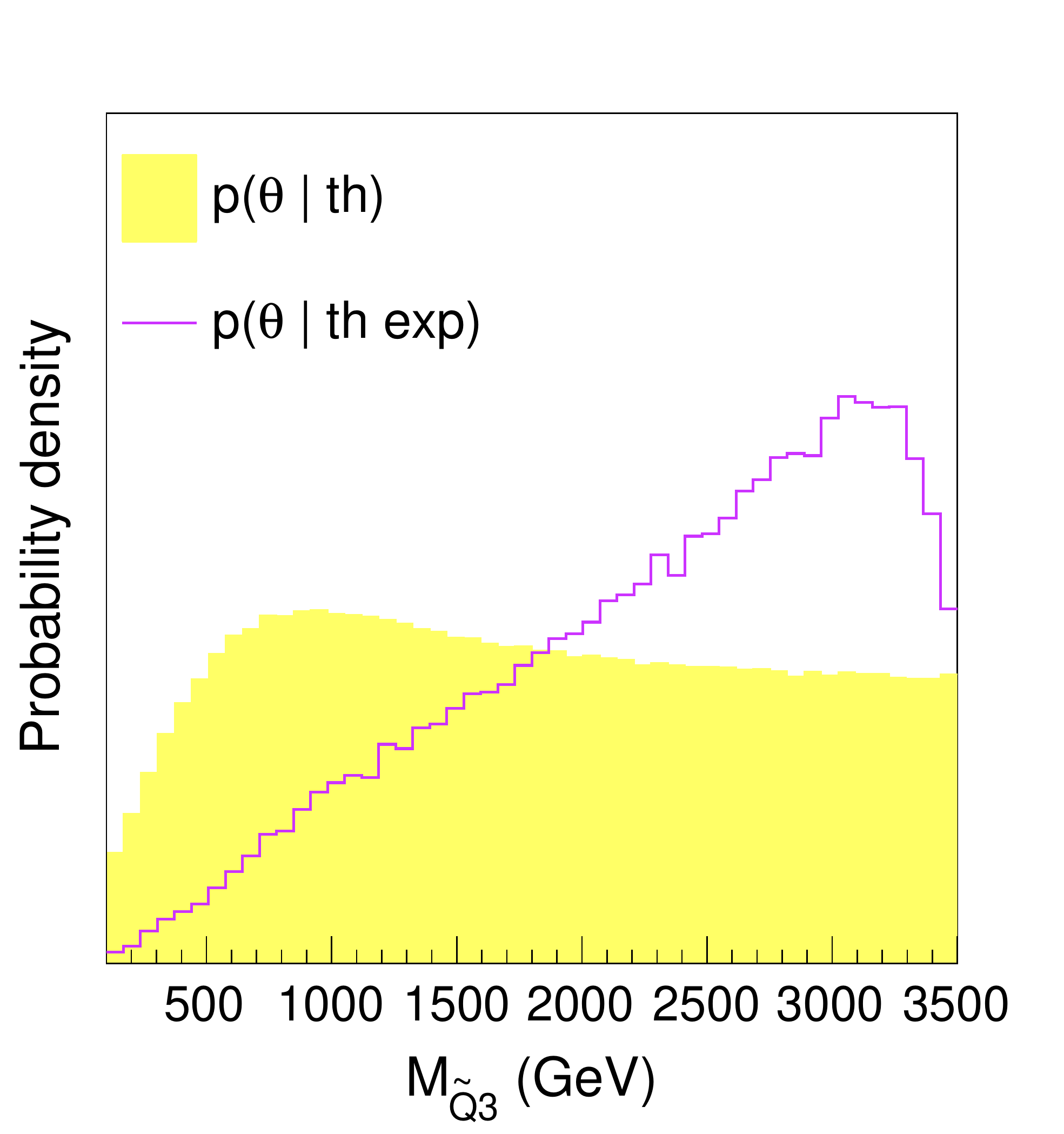} 
		\includegraphics[width=0.32\textwidth]{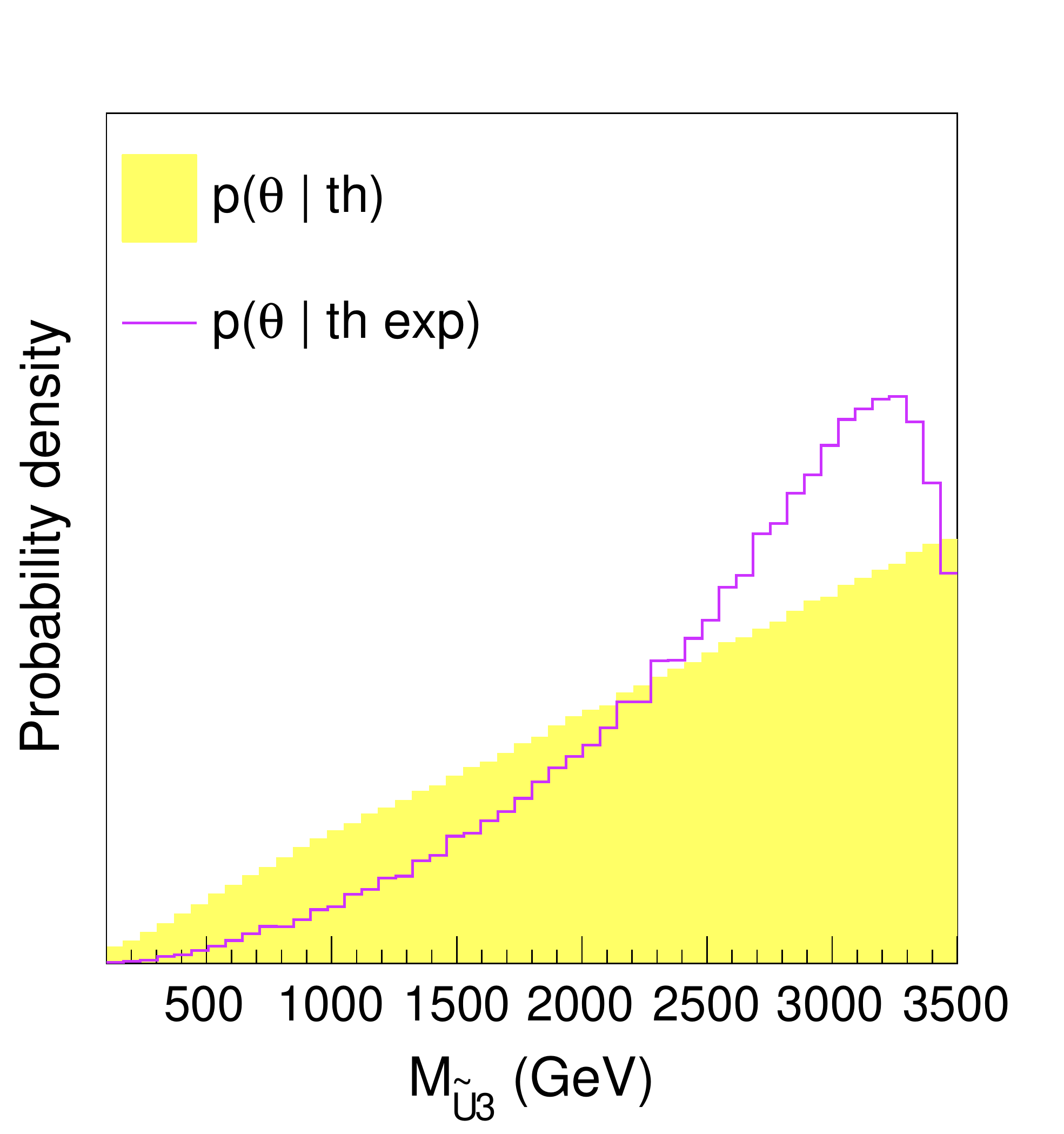} 
	\end{center}
	\vspace*{-8mm}
	\caption{The one-dimensional prior (yellow histogram) and posterior
  (violet curve) distributions of the parameters of our NMFV MSSM description. The
  prior only incorporates theoretical inputs while the posterior distribution
  shows the impact of all experimental observations listed in Table 1.}
	\label{FigQFVdist1}
\end{figure}

Figure \ref{FigQFVdist1} shows the prior distribution, obtained from a uniform random scan and excluding only non-physical parameter configurations, together with the posterior distribution, obtained from the MCMC scan imposing all mentioned constraints. The bino mass is preferred to be rather low. For the trilinear coupling $A_f$, the posterior distribution features two sharp peaks at $|A_f| \sim 3000$ GeV. This comes from the Higgs mass condition $m_h \sim 125$ GeV, which favours a large stop mass splitting. More precisely, defining $X_t = (A_u)_{33} - \mu/\tan\beta$ and $M^2_{\rm SUSY} = m_{\tilde{q}_1} m_{\tilde{q}_2}$ with $\tilde{q}_{1,2}$ being the two squarks exhibiting the largest stop contents, the leading contributions to $m_h$ can be expressed as
\begin{equation}
	m^2_h ~=~ m^2_Z \cos^2 2\beta + \frac{3 g^2 m_t^4}{8\pi m^2_W} \left[ \log\frac{M^2_{\rm SUSY}}{m^2_t} 
		+ \frac{X_t^2}{M^2_{\rm SUSY}} \left( 1 - \frac{X_t^2}{12 M^2_{\rm SUSY}} \right) \right] .
	\label{Eq:HiggsMass1}
\end{equation}
Therefore, imposing $m_h \sim 125$ GeV leads to $|X_t| \sim \sqrt{6} M_{\rm SUSY}$ corresponding to the observed peaks in $A_f$. 

For the squark soft mass parameters, large values are preferred for the third generation, while the parameters associated to the first and second generations are pushed towards the lower end of the scanned interval. This somewhat surprising behaviour is also related to the Higgs mass. In the simplified case where all squarks have a similar mass $\tilde{m}$, the one-loop corrections including left-right mixing can be approximated by \cite{Kowalska:2014opa}
\begin{equation}
	\Delta m^2_h = \frac{3 v_u^4}{8 \pi^2(v^2_d+v^2_u)} 
	\left[ \frac{(T_u)_{23}^2}{\tilde{m}^2} \left( \frac{(Y_u)^2_{33}}{2} - \frac{(T_u)_{23}^2}{12 \tilde{m}^2} \right)\right] ,
	\label{Eq:HiggsMass2}
\end{equation}
where in our parametrization $(T_u)_{23} \propto \delta^u_{LR}$ (see Eq.\ (\ref{Eq:NMFVParams})). The flavour constraints push the third generation masses to higher values, such that only relatively low values of $M_{\tilde{Q}_{1,2}}$ can maintain a large enough NMFV parameter $\delta^u_{LR}$ to make the required significant contribution to $m_h$. Similar arguments hold for the other parameters including the sector of down-type squarks. 

\begin{figure}[t]
	\begin{center}
		\includegraphics[width=0.32\textwidth]{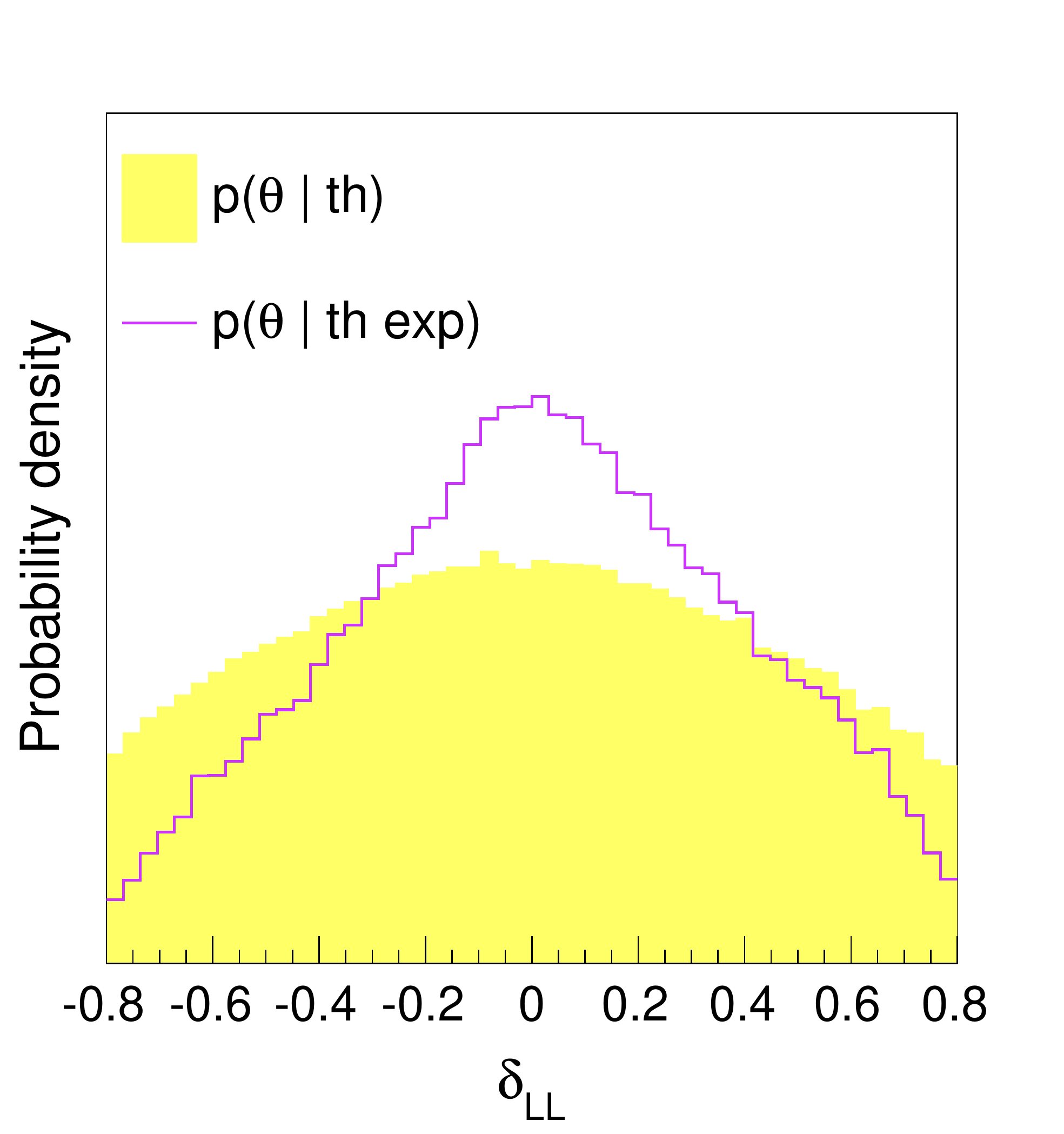} 
		\includegraphics[width=0.32\textwidth]{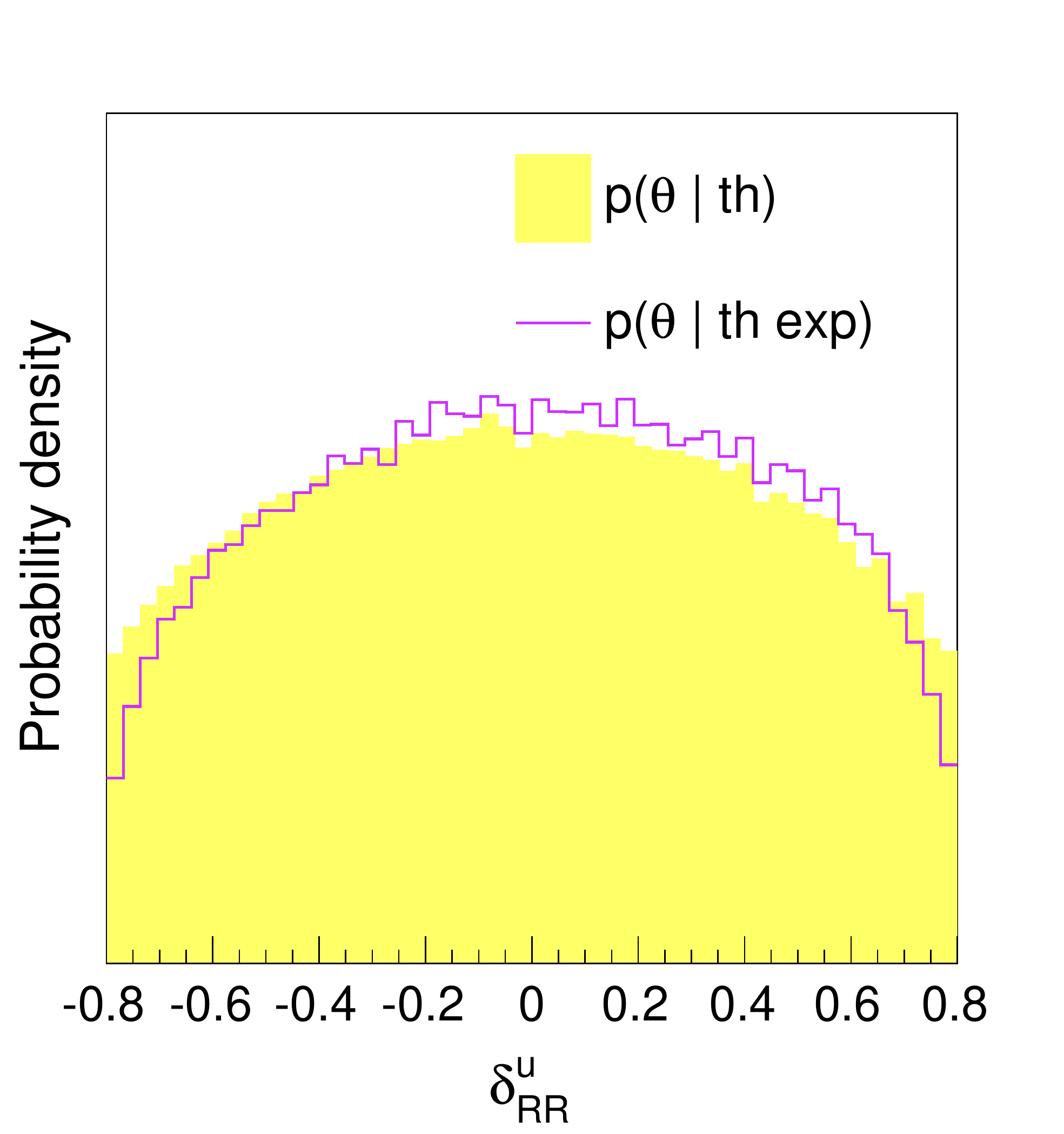} 
		\includegraphics[width=0.32\textwidth]{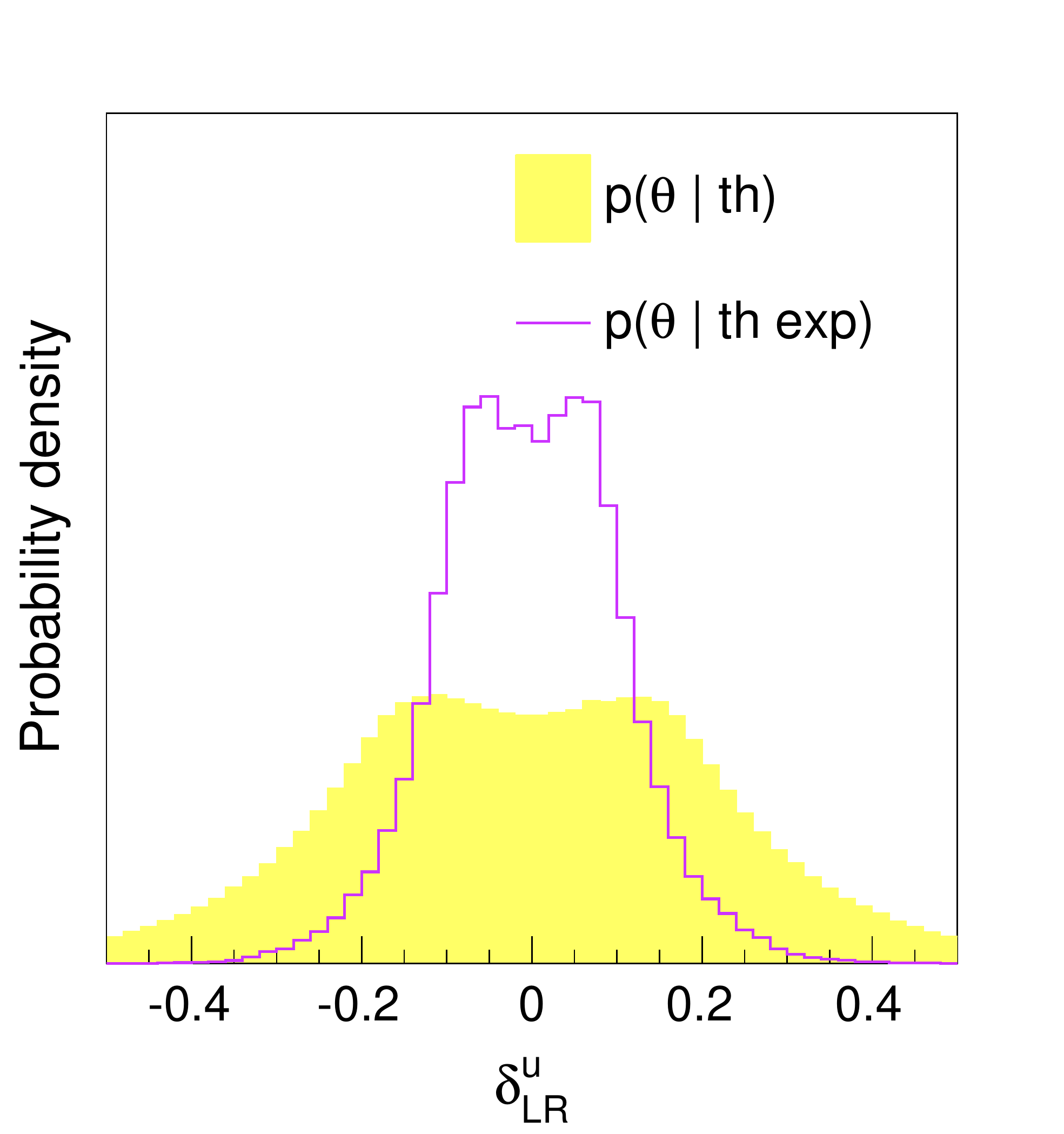} 
	\end{center}
	\vspace*{-8mm}
	\caption{Same as Figure 1 for three selected QFV input parameters.}
	\label{FigQFVdist2}
\end{figure}

Turning to the QFV terms, we show the distributions of three selected NMFV parameters in Figure \ref{FigQFVdist2}. As can be seen, left-left and right-right mixing is only mildly constrained, such that the corresponding NMFV parameters can be rather large ($|\delta_{LL}| \lesssim 0.8$ and $|\delta^u_{RR}| \lesssim 0.8$). Left-right and right-left mixing is mainly constrained by the Higgs boson mass. The distributions of $\delta^u_{LR}$ exhibits two peaks, which are expected again from Eq.\ (\ref{Eq:HiggsMass2}) and the fact that $(T_u)_{23} \propto \delta^u_{LR}$. An analysis of the global situation shows that a large fraction of the scanned points exhibits seven non-vanishing NMFV parameters, out of which some are numerically sizeable (see Ref.\ \cite{OurPaper}).

\begin{figure}[t]
	\begin{center}
		\includegraphics[width=0.32\textwidth]{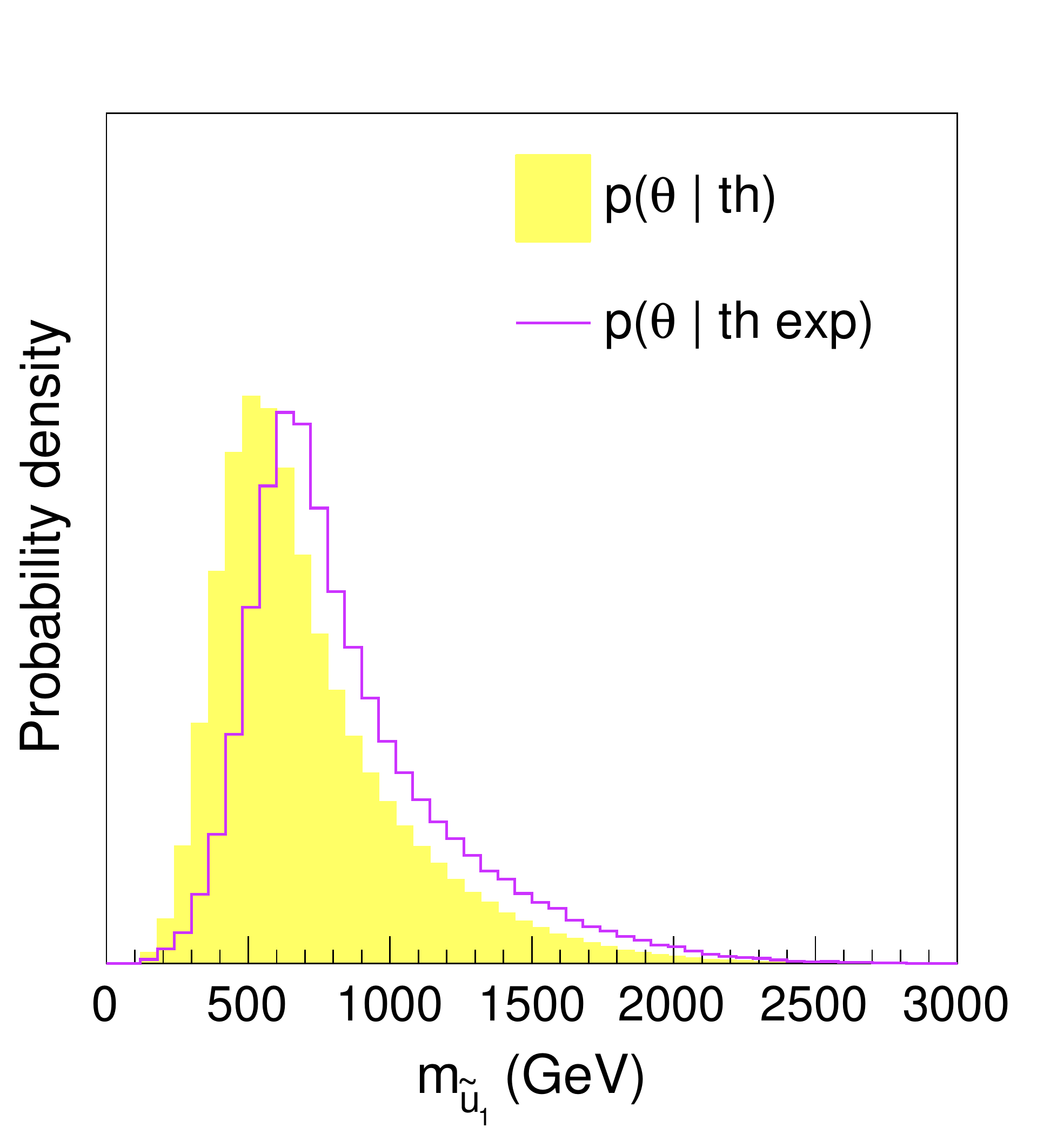} 
		\includegraphics[width=0.32\textwidth]{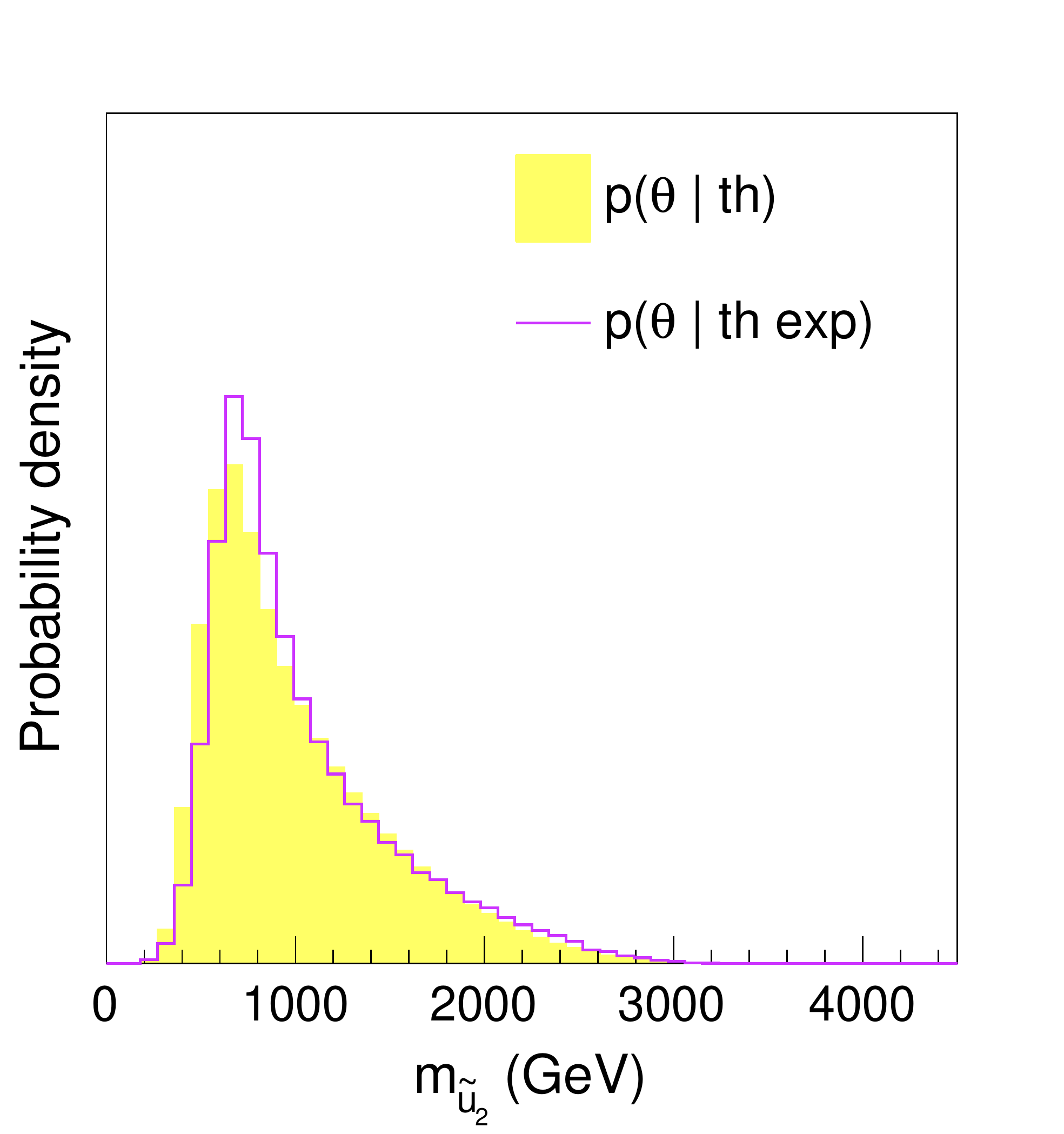} 
		\includegraphics[width=0.32\textwidth]{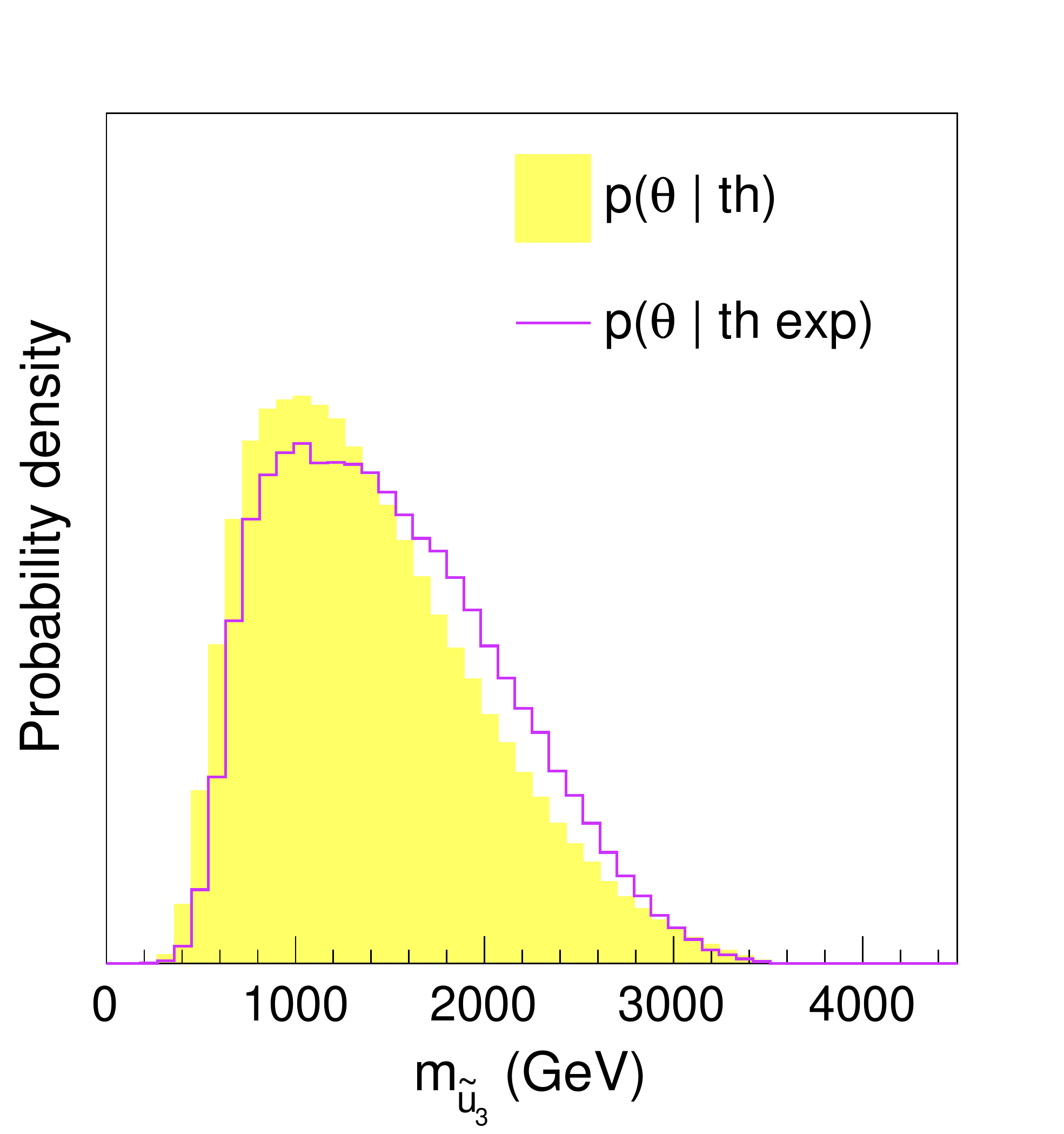} 
	\end{center}
	\vspace*{-8mm}
	\caption{One-dimensional prior (yellow histogram) and posterior (violet curve) distributions of the masses of three lightest up-type squarks.}
	\label{FigMassdist1}
\end{figure}

\begin{figure}[t]
	\vspace*{5mm}
	\begin{center}
		\includegraphics[width=0.45\textwidth]{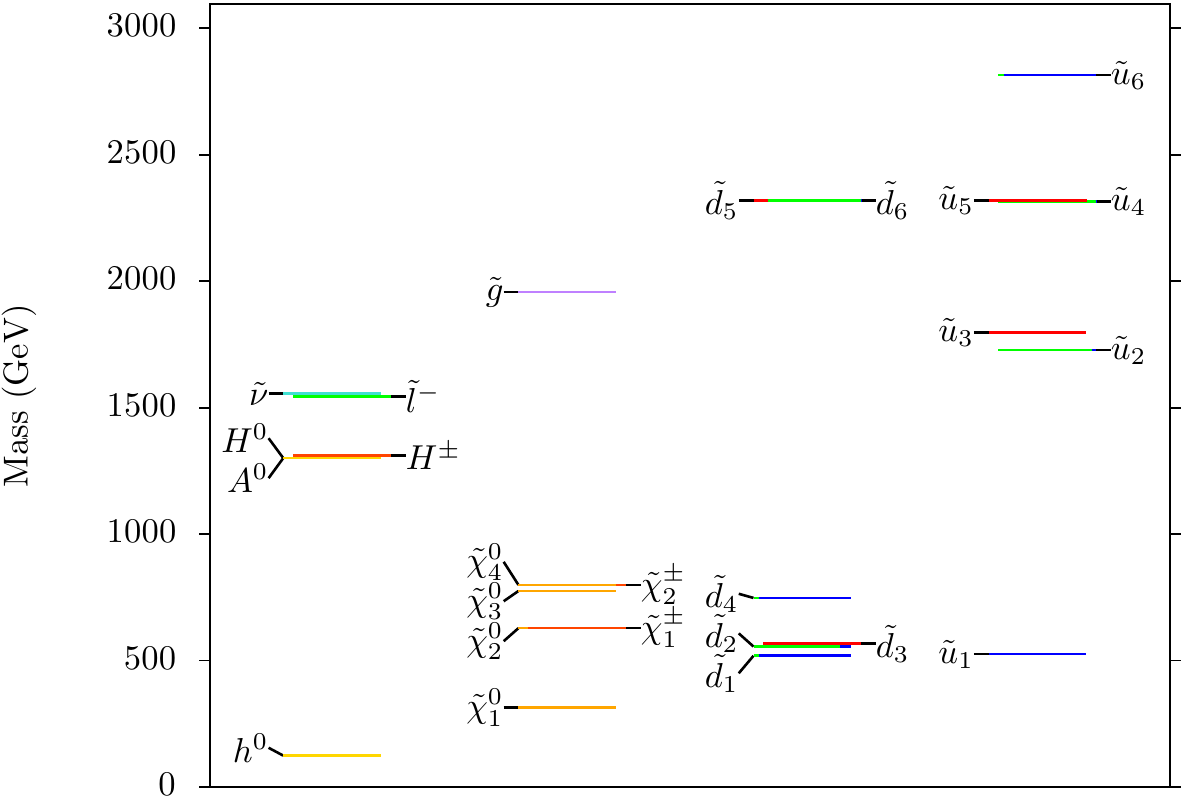}
		\includegraphics[width=0.45\textwidth]{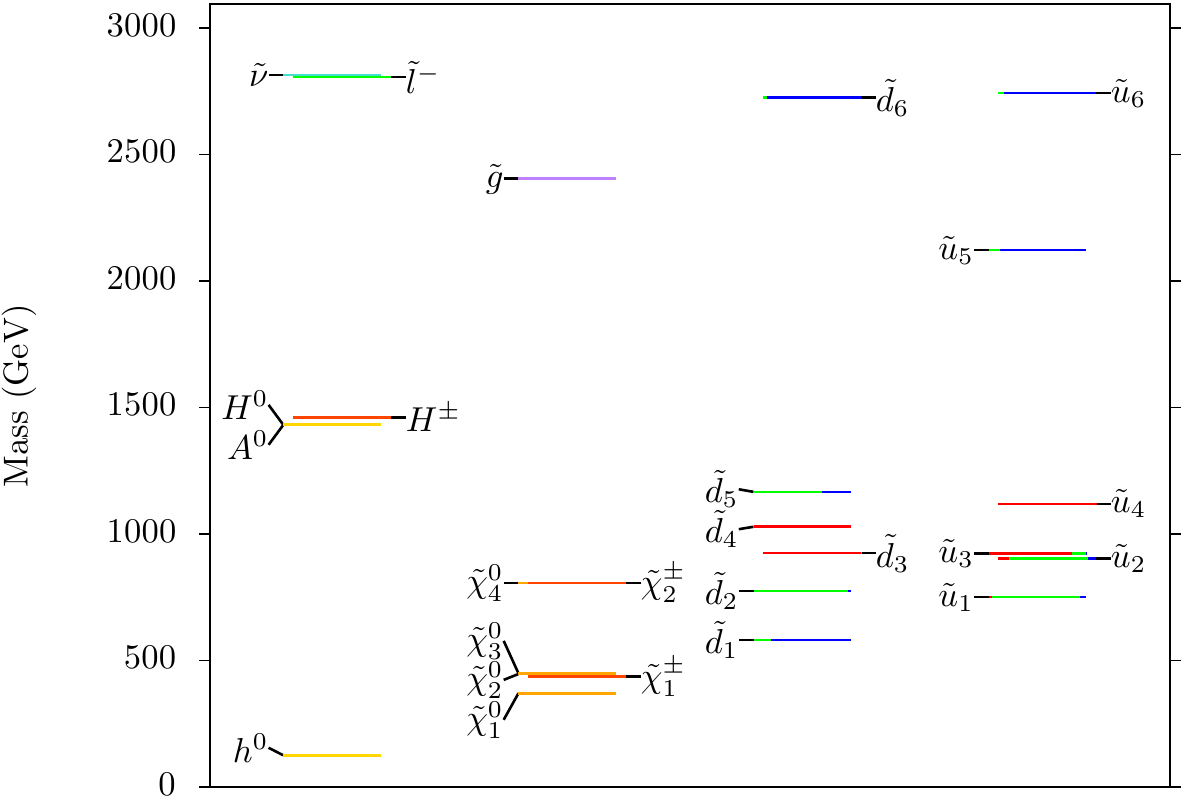}
	\end{center}
	\vspace*{-4mm}
	\caption{Mass spectra for two example benchmark scenarios. For the squarks, the colour code indicates the flavour content: \{red, green, blue\} corresponds to \{first, second, third\} generation flavours.}
	\label{FigBenchmarks}
\end{figure}

Let us now turn to the resulting physical squark eigenstates. In Figure \ref{FigMassdist1} we show the prior and posterior distributions of the masses of the three lightest up-type squarks for selected QFC parameters. The two lightest states are generally relatively close in mass. It is interesting to note that they contain mostly first and second generation flavour (see Figure \ref{FigBenchmarks} for two example scenarios), while in MFV MSSM scenarios the lightest squarks are often the stops. This difference is due to the fact that the first and second generation soft masses are pushed towards lower values as explained above.

For most viable points, the mass difference between the lightest squark and the lightest neutralino is well above 50 GeV, which is a favorable condition for collider searches. A considerable number (40\%) of parameter points features $\tilde{u}_1$ masses of about 500 -- 1000 GeV together with neutralino masses of the order of 150 -- 400 GeV. Note that such mass configurations are likely to be ruled out by recent LHC data.

Taking into account this last constraint, we finally define benchmark points to set the stage for future studies. The scenarios are chosen to capture the different phenomenological properties revealed by our MCMC analysis. In Figure \ref{FigBenchmarks} we show the mass spectra of two benchmark scenarios. The proposed scenario feature several squark eigenstates below 1 TeV, i.e.\ reachable at the LHC. In addition, large flavour mixing between the second and third generations is present for those states leading to a rich phenomenology thanks to a large variety of open decay modes. 

\section{Conclusion}

We have presented a Markov Chain Monte Carlo study of a 19-parameter TeV-scale MSSM including NMFV terms corresponding to mixing between the second and third generation squarks. We have explored the parameter space with respect to a set of theoretical and experimental constraints. The latter include limits arising from $B$-meson and kaon physics as well as the Higgs boson mass of about 125 GeV. 

Our study has shown that in the NMFV MSSM, the lightest squarks are often not stop- or sbottom-like, which contrasts with scenarios of the usual minimally flavour-violating MSSM. Requiring a theoretically consistent Higgs sector and a light Higgs boson of about 125~GeV similarly restrict the left-right and right-left flavour-violating squark mixing parameters $\delta_{LR/RL}^{u,d}$ to be small. In contrast, the $\delta_{LL}$ and $\delta_{RR}^d$ NMFV parameters are mainly constrained by neutral $B$-meson oscillations and rare $B_s \rightarrow \mu \mu$ decay mainly influences $\delta_{RL}^u$. All other NMFV parameters are left unconstrained by the considered experimental observations.

Based on the results of the MCMC scan, we have defined four benchmark scenarios allowed by current data and exhibiting distinct features. In each scenario, several squarks have masses below or close to 1 TeV such that they will be reachable within the next few years. These scenarios are suitable for future analyses of NMFV effects in the MSSM at colliders. 

For a more detailed discussion of all mentioned aspects the reader is referred to Ref.\ \cite{OurPaper}.


\end{document}